# Patient-specific mean teacher UNet for enhancing PET image and low-dose PET reconstruction on RefleXion X1 biology-guided radiotherapy system

Jie Fu, PhD[1], Zhicheng Zhang, PhD[1], Linxi Shi, PhD[2], Zhiqiang Hu, PhD[2], Thomas Laurence, PhD[2], Eric Nguyen, BS[1], Peng Dong, PhD[1], Guillem Pratx, PhD[1], Lucas Vitzthum, MD[1], Daniel T. Chang, MD[1], Lei Xing, PhD[1], and Wu Liu, PhD[1*]

1. Department of Radiation Oncology, Stanford University, Stanford, CA, 94305

2. RefleXion Medical, Inc., Hayward, CA, 94545

*Email: wuliu@stanford.edu

## Abstract

**Background:** The RefleXion® X1 is the first biology-guided radiotherapy (BgRT) system combining PET imaging and fan beam CT with a 6 MV flattening-filter-free linear accelerator. Its dual 90-degree PET detector provides enough space for the radiation delivery system. However, due to its reduced geometric efficiency, RefleXion PET collects fewer pair production events compared to a full-ring diagnostic PET system. In the proposed BgRT workflow, a short scan is acquired before treatment delivery to ensure image quality and consistency. The shorter scan time, a quarter of the simulation scan time, also leads to fewer coincidence events and hence reduced image quality.

**Purpose:** In this study, we proposed a patient-specific mean teacher UNet (MT-UNet) to enhance PET image quality and low-dose PET reconstruction on RefleXion X1.

**Methods:** PET/CT scans of nine cancer patients were acquired using RefleXion X1. Every patient had one simulation scan. Five patients had additional scans acquired during the first and the final treatment fractions. Treatment scans were acquired using the same imaging protocol as the simulation scan. Detected coincidence events were saved in list-mode data format. For each scan, we reconstructed a full-dose image and evenly split coincidence events into four sessions to reconstruct four quarter-dose PET images. For each patient, our proposed MT-UNet was trained using quarter-dose and full-dose images of the simulation scan. MT-UNet contains a student UNet and a teacher UNet. Parameters of the student UNet were optimized during training, while parameters of the teacher UNet were kept the same as moving-average values of the student UNet parameters. Consistency loss between the outputs of teacher and student UNets was included to train the MT-UNet. For the image quality enhancement task, we applied nine trained MT-UNets to full-dose simulation PET images of the nine patients to generate enhanced images, respectively. The enhanced images were compared with the original full-dose images using the contrast-to-noise ratio (CNR) and signal-to-noise ratio (SNR). For the low-dose image reconstruction task, we applied five trained MT-UNets to ten quarter-dose treatment images of five patients to predict full-dose images, respectively. The predicted and ground truth full-dose images were compared using structural similarity index measure (SSIM) and peak signal-to-noise ratio (PSNR). We also trained and evaluated patient-specific UNets for model comparison.

**Results:** PET images enhanced by the MT-UNet achieved the highest average CNR and SNR compared with full-dose simulation PET images and those enhanced by the UNet. Images predicted by the MT-UNet achieved the highest average SSIM and PSNR compared with quarter-dose treatment images and those predicted by the UNet. The *p*-values of Wilcoxon signed-rank tests for SSIM and PSNR between the UNet and MT-UNet were less than 0.05.

**Conclusion:** Our proposed patient-specific MT-UNet achieved better performance in improving the quality of RefleXion low-dose and full-dose images compared to the patient-specific UNet. The MT-UNet demonstrated potential in improving the quality of the short PET setup scan before treatment delivery and early prediction of the full treatment scan shortly after the treatment starts. Our novel training strategies can also be adapted to diagnostic PET systems.

# 1. INTRODUCTION

The RefleXion® X1 (RefleXion Medical, Inc., Hayward, CA) is the first biology-guided radiotherapy (BgRT) system combining dual imaging technologies (PET and CT) with a 6 MV flattening-filter-free linear accelerator[1]. The RefleXion X1 has dual 90-degree PET detector arcs designed to acquire PET emissions for guiding treatment delivery during the BgRT[1,2]. This is different from most diagnostic PET imaging systems that have a full ring of PET detectors. While the PET scan acquired on the X1 supports radiation delivery, as expected, the resultant PET images have lower image quality compared to diagnostic PET images. Improved PET image quality could potentially enable more accurate tumor delineation and tracking. Therefore, we are aiming to improve the quality of the RefleXion PET images.

The radiation dose from PET scans to patients and technicians would theoretically result in an increased risk of cancer[3–5]. Reducing injected tracer dose could lead to less radiation dose but also poor image quality if using conventional image reconstruction algorithms. Additionally, the proposed BgRT workflow requires the acquisition of a short PET scan before treatment delivery to ensure image quality and consistency. Short scan time is preferred to avoid possible large patient motions during the scan. The proposed scan time is a quarter of the simulation scan time. This leads to low-quality "quarter-dose" images because of the reduced number of coincidence events. Also, RefleXion X1 allows the acquisition of PET images during the treatment. Currently, both simulation PET scan and treatment PET scan share the same imaging protocol in which patients on the couch would pass back and forth four times. There is an increasing desire for displaying high-quality images after each treatment pass to verify the treated anatomy. Similar to the short scan, the images acquired after each pass or the beginning few passes would have low image quality as "quarter-dose" images. Our additional goal is to improve the quality of the low-dose RefleXion PET images.

Recently, many deep learning[6] models have achieved promising performance in improving low-dose PET reconstruction. Ouyang et al. proposed a generative adversarial network (cGAN) with feature matching and perceptual loss to generate high-quality brain PET images from ultra-low-dose images[7]. Wang et al. investigated the concatenated conditional GANs to improve the quality of low-dose Brain PET images[8,9]. Zhou et al. used a cycle Wasserstein regression adversarial training framework to denoise low-dose lung PET images[10,11]. The generators in these models all have UNet-like architectures[12]. However, these models were only investigated for diagnostic PET images which have better image quality compared to those acquired with RefleXion X1. Additionally, the authors only investigated the population-based models that were trained using images from many patients. Training population-based models have several limitations. First, it requires a large training dataset which may be challenging to acquire. Second, each anatomy site may require training a site-specific model for good generalization. Finally, the model may not be robust when applied to new patients[13,14].

To address those limitations, we want to investigate the feasibility of training patient-specific deep learning models to enhance the quality of full-dose RefleXion X1 PET images and low-dose RefleXion PET reconstruction. PET images are normally reconstructed using coincidence events, so we can generate many low-dose PET images by sampling coincidence events. This allows us to generate more images for training the patient-specific model. Also, the mean teacher technique proposed by Tarvainen and Valpola helped achieve better semi-supervised deep learning results[15]. In this study, we proposed a patient-specific mean teacher UNet (MT-UNet) that can be trained using four low-dose images of one PET scan.

## 2. MATERIALS AND METHODS

### 2. A. Dataset

PET/CT scans of nine lung cancer and bone metastases patients were acquired with the RefleXion X1 machine using 15 mCi of (18)F-fluorodeoxyglucose. Cancer sites include lung (n=4), pelvis (n=2), chest wall (n=1), spine (n=1), rib (n=1), and hip (n=1). Every patient had one simulation PET/CT scan. Five out of nine patients also had two additional treatment PET/CT scans acquired, during the first and the final treatment fractions, using the same imaging protocol as the simulation scan. It should be noted that the patients were treated using stereotactic body radiation therapy, not BgRT. Detected coincidence events of each PET scan were saved in list-mode data format. We evenly split coincidence events into four sessions to reconstruct four quarter-dose PET images which were then post-filtered using RefleXion clinical standard parameters. Similarly, we also reconstructed full-dose PET images using all coincidence events. The top and bottom four slices were removed due to the intrinsic high noise levels. The in-plane dimension of PET images is 128 x 128 with a pixel size of 4 x 4 mm$^2$. The slice thickness is 2.1 mm.

We generated a body mask from the CT image for every PET/CT scan using Otsu's thresholding[16] followed by opening and closing morphological operations. PET images were normalized using the mean intensity of voxels within the body mask. The intensity of voxels outside the body was set to 0. Figure 1 shows the transverse slices of four quarter-dose PET images and a full-dose PET image of one simulation PET scan. As expected, quarter-dose images have lower image quality than full-dose images.

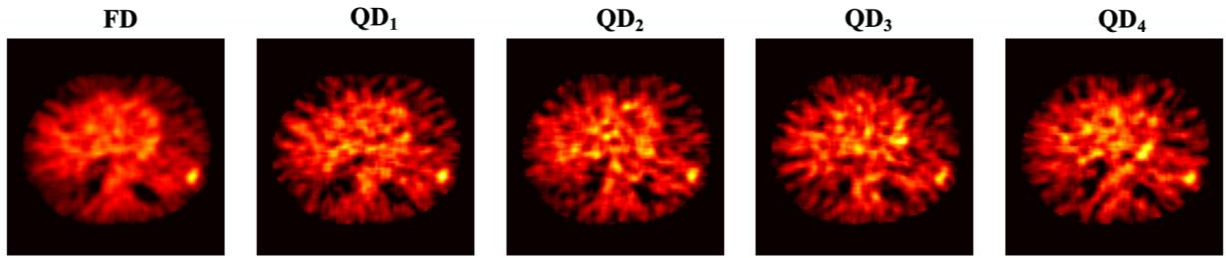

Figure 1. Transverse slices of full-dose (FD) PET images and four quarter-dose (QD) PET images of one simulation PET scan.

## 2. B. Patient-specific UNets

Our proposed patient-specific MT-UNet contains two identical UNets, a student UNet (S-UNet) and a teacher UNet (T-UNet), as shown in Figure 2. For each patient, we trained an MT-UNet using quarter-dose and full-dose images of the simulation scan. A UNet was also trained for model comparison.

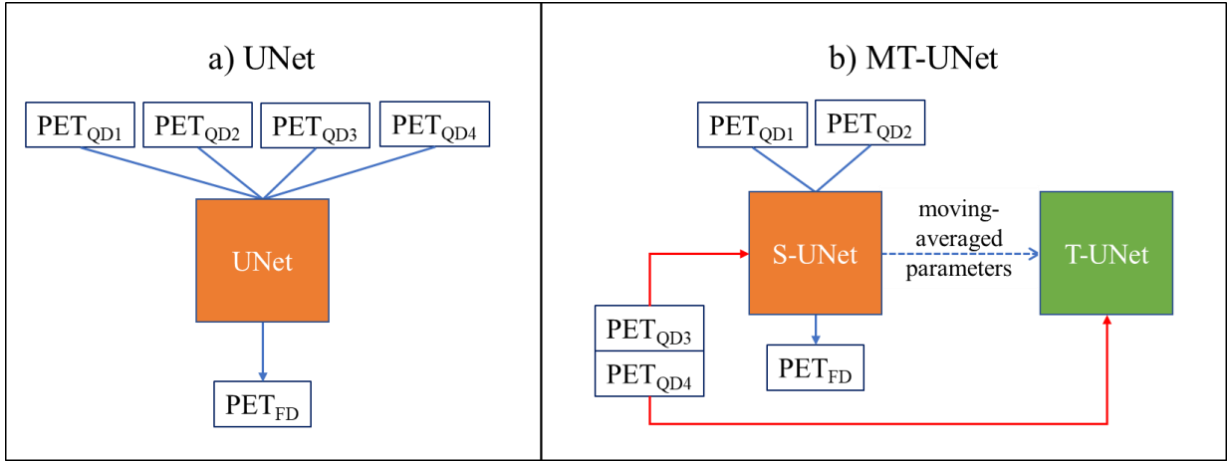

Figure 2. Training scheme of the UNet and MT-UNet. QD, quarter dose; FD, full dose; S-UNet, student-UNet; T-UNet, teacher-UNet.

As shown in Figure 2 a), four quarter-dose images were input into the UNet to predict the full-dose image. The UNet parameters were then optimized by minimizing the sum of mean squared errors (MSEs) between four predicted images and ground truth full-dose images, i.e. image reconstruction loss,

$$L_{UNet} = \sum_{i=1}^{4} MSE(UNet(PET_{QDi}), PET_{FD}),$$

(1)

where $PET_{QDi}$ is the $i^{th}$ quarter-dose PET image; $UNet(PET_{QDi})$ is the output image of the UNet based on $PET_{QDi}$; $PET_{FD}$ is the full-dose PET image.

As shown in Figure 2 b), two quarter-dose images were input into the S-UNet of MT-UNet to compute the image reconstruction loss. Additionally, the other two quarter-dose images were input into both the S-UNet and T-UNet to compute the prediction consistency loss, which equals the sum of MSEs between the images predicted by the S-Unet and the images predicted by the T-UNet;

$$L_{MT-Net} = \sum_{i=1}^{2} MSE(S\text{-}UNet(PET_{QDi}), PET_{FD})$$
$$+ \lambda \times (\sum_{i=3}^{4} MSE(S\text{-}UNet(PET_{QDi}), T\text{-}UNet(PET_{QDi}))),$$

(2)

where $S\text{-}UNet(PET_{QDi})$ is the output image of the S-UNet based on $PET_{QDi}$; $T\text{-}UNet(PET_{QDi})$ is the output image of the T-UNet based on $PET_{QDi}$; $\lambda$ is the regularization parameter. During the training, the S-UNet parameters were optimized by minimizing both the image reconstruction loss and prediction consistency loss, while the T-UNet parameters were kept the same as exponential moving-average values of the S-UNet parameters.

Both UNets were implemented using the Tensorflow package (V1.10.0, Python 3.6.9, CUDA 10.1). A batch size of 16 was used for training. The Adam stochastic gradient descent method[17] was used to minimize the loss function. The initial learning rate and the exponential moving average decay value were set as $5 \times 10^{-4}$ and 0.995, respectively. Loss regularization parameter, $\lambda$, was set as 0.05. Both models were trained for 1200 epochs. The model resulting in the smallest loss was saved for PET image quality enhancement and low-dose prediction.

## 2. C. PET image quality enhancement

We investigated the feasibility of using patient-specific MT-UNet to enhance the full-dose PET image. For every patient, the full-dose PET image of the simulation scan was input into the trained UNet and MT-UNet to generate the enhanced PET image. The performance of image quality enhancement was evaluated using contrast-to-noise (CNR) and signal-to-noise ratio (SNR),

$$CNR = \frac{\mu_{tumor} - \mu_{background}}{\sigma_{background}} \text{ and } SNR = \frac{\mu_{tumor}}{\sigma_{background}},$$

(3)

where $\mu_{tumor}$ and $\mu_{background}$ are the mean values of voxels within the tumor region-of-interest (ROI) and the background ROI, respectively; $\sigma_{background}$ is the standard deviation of voxel within the background ROI.

## 2.D. Low-dose PET reconstruction

We also investigated the feasibility of using patient-specific MT-UNet to predict the full-dose PET image from the quarter-dose image. For five patients who had two treatment PET/CT scans, the quarter-dose PET image of each treatment scan was input into the trained UNet and MT-UNet to predict full-dose PET images. As both models were trained only using simulation images, treatment images were unseen to the models. The performance of the low-dose PET reconstruction was evaluated by comparing predicted and ground truth full-dose treatment images using structural similarity index measure (SSIM) and peak signal-to-noise ratio (PSNR),

$$SSIM(x,y) = \frac{(2\mu_x\mu_y+C_1)(2\sigma_{xy}+C_2)}{(\mu_x^2+\mu_y^2+C_1)(\sigma_x^2+\sigma_y^2+C_2)} \text{ and } PSNR(x,y) = 20 \times log_{10}(\frac{MAX(y)}{\sqrt{MSE(x,y)}})$$

(4)

where x and y represent the signals of the predicted image and ground truth images, respectively; $\mu_x$ and $\mu_y$ represent the mean intensity of x and y, respectively; $\sigma_x$ and $\sigma_y$ represent the variance of x and y, respectively; $\sigma_{xy}$ represent the covariance of x and y; $C_1$ and $C_2$ are constant stabilizing terms. SSIM and PSNR were computed using central 20 slices.

## 3. RESULTS

### 3. A. PET image quality enhancement

Figure 3 shows the transverse slices of the original full-dose PET image and model-enhanced PET images for the two example patients along with tumor and background ROIs. Table 1 summarizes the statistics of CNR and SNR for tumors in PET images. The PET images enhanced by the MT-UNet achieved the highest average CNR and SNR.

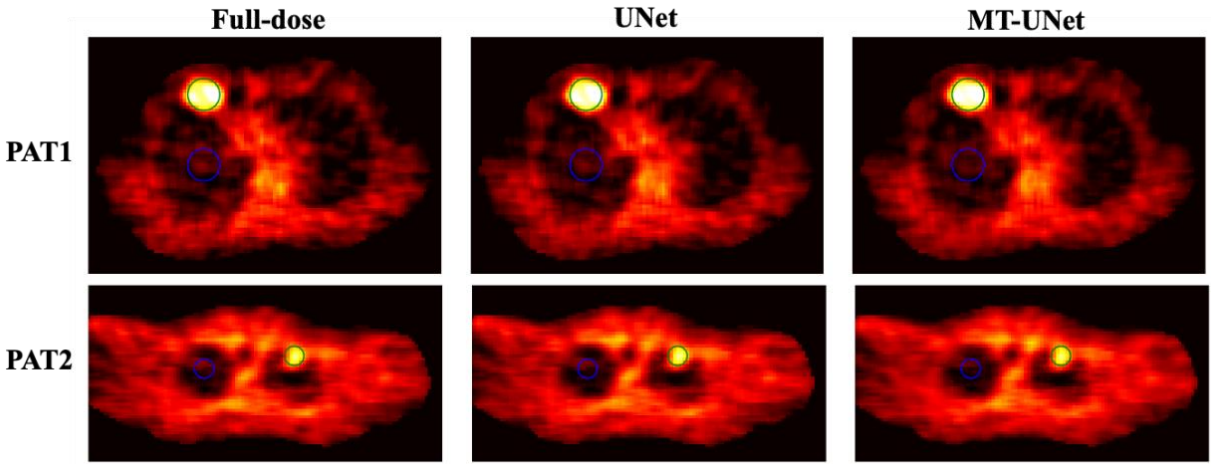

Figure 3. Transverse slices from the original full-dose PET images and the PET images enhanced by the UNet and MT-UNet for two lung cancer patients. Green circles and blue circles show tumor ROIs and lung ROIs, respectively.

|     | Full-dose | UNet | MT-UNet |
| --- | --- | --- | --- |
| **CNR** | 12.06 ± 5.12 | 13.13 ± 5.65 | 15.52 ± 7.57 |
| **SNR** | 15.83 ± 4.16 | 16.81 ± 4.66 | 19.83 ± 6.86 |

Table 1. Statistics of CNR and SNR of the tumors in the original full-dose PET, UNet-enhanced PET, and MT-UNet-enhanced PET images. Results were averaged across 9 patients and shown in (mean ± SD) format.

### 3.B. Low-dose PET reconstruction

Figure 4 shows the transverse slices of treatment PET images from the first and the final treatment fractions and the corresponding PET images predicted by the UNet and MT-UNet for one example patient. Both models were trained using one simulation PET scan. Both models generated higher quality PET images compared to low-dose images based on visual inspections. Table 2 summarizes the statistics of SSIM and PSNR between predicted images and high-dose images. The PET images reconstructed using the MT-UNet achieved the highest average SSIM and PSNR. The *p*-values of Wilcoxon signed-rank tests for metrics between the UNet and MT-UNet were less than 0.05, which indicates strong evidence against the null hypothesis that the median difference between the paired samples is 0.

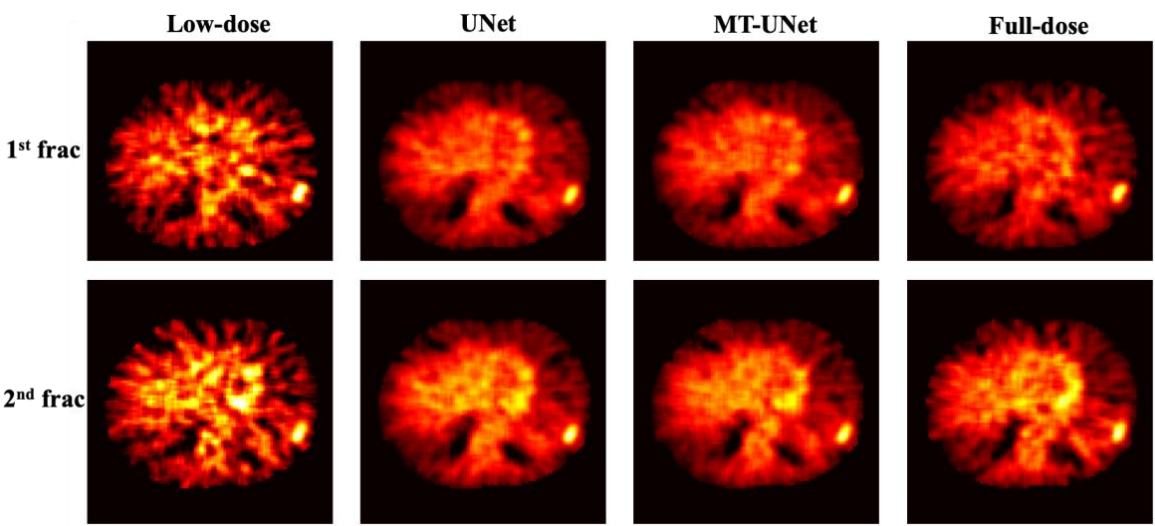

Figure 4. Transverse slices from full-dose PET images of the first and the final treatment fractions along with the corresponding low-dose images and the images predicted by the UNet and MT-UNet for one example patient.

|  | **Low-dose** | **UNet** | **MT-UNet** |
|---|---|---|---|
| **SSIM** | 0.912 ± 0.037 | 0.935 ± 0.028 | 0.940 ± 0.027** |
| **PSNR [dB]** | 30.854 ± 4.319 | 32.326 ± 3.429 | 32.747 ± 3.486** |

Table 2. Statistics of SSIM and PSNR of the low-dose image and the images predicted by the UNet and MT-UNet. Results were averaged across 10 treatment images. **The *p*-values of Wilcoxon signed-rank tests for metrics between the UNet and MT-UNet were less than 0.05

### 4. DISCUSSION

In this study, we proposed a novel patient-specific MT-UNet to enhance PET image quality and low-dose PET reconstruction for the RefleXion X1 machine. To our knowledge, this is the first study that applied patient-specific deep learning models to denoise low-dose and full-dose PET images. Sampling coincidence events allows us to generate many low-dose PET images from only one simulation scan, which make it possible to train a patient-specific model. Additionally, integrating the mean teacher technique into training a UNet, i.e. MT-UNet, helped generate higher quality PET images in both tasks compared to the conventional UNet as indicated by Tables 1 and 2.

The patient-specific MT-UNet improved the image quality of low-dose RefleXion PET images. Low-dose reconstruction allows us to reduce scan acquisition time. We mentioned that a short (1/4) PET scan is acquired before treatment delivery in the proposed BgRT workflow. During the treatment, PET images could also be displayed after each pass of 4 treatment passes to verify the treated location. Both short set-up scans and scans after each pass have similar image quality as the quarter-dose image. However, quarter-dose images have high noise levels as shown in Figure 4. Our study demonstrated the potential of using the MT-UNet to improve the quality of short scans and scans after each treatment pass. Additionally, the MT-UNet can improve the image quality of full-dose RefleXion images. Higher quality images could potentially lead to more accurate tumor contours. Future studies are needed to investigate the performance of using model-enhanced PET images for tumor contour.

Our proposed model can also be trained for low-dose reconstruction in diagnostic PET systems. Unlike the PET arcs of RefleXion X1, diagnostic PET systems have higher geometry efficiency which could result in images with higher SNR. This could be beneficial for training an accurate model as both the input low-dose images and ground truth full-dose images have better image quality. Future work includes acquiring diagnostic PET scans and adapting the MT-UNet for PET image quality enhancement and low-dose PET reconstruction for diagnostic PET systems.

Our study demonstrated the feasibility of training patient-specific models for enhancing both the low-dose and high-dose PET images in multiple anatomy sites such as lung and pelvis. Training population-based models for multiple anatomy sites requires a large number of PET images from these sites. The main advantage of training patient-specific models is reducing the need to acquire a large training dataset. It should be noted that mean teacher and PET image sampling techniques investigated in this study can also be adapted to train population-based models. In the future study, we will include more patients to compare the patient-specific MT-UNet with population-based deep learning models and enhance the real-time PET acquisition for motion tracking performance improvement.

## 5. CONCLUSION

Our proposed patient-specific MT-UNet achieved promising performance in improving PET image and low-dose PET reconstruction for RefleXion X1. MT-UNet achieved better performance in both tasks compared to UNet. Our novel training strategies can also be adapted to diagnostic PET systems and population-based deep learning models.

**ACKNOWLEDGMENT**

The authors would like to thank RefleXion Medical, Inc. for the research support and Camellia Djebroun for collecting patient treatment information.

**CONFLICTS OF INTEREST**

- Linxi Shi, Zhiqiang Hu, and Thomas Laurence are employees of RefleXion Medical Inc.
- Wu Liu, Lucas Vitzthum, and Daniel T. Chang have received research support from RefleXion Medical, Inc.